\documentclass[twocolumn,3p,times]{elsarticle}

\usepackage[utf8]{inputenc}
\usepackage[T1]{fontenc}
\usepackage[english]{babel}
\usepackage{graphicx}
\usepackage{exscale}
\usepackage{color}
\definecolor{grey}{gray}{0.4}
\usepackage{setspace}
\usepackage{amsmath}
\usepackage{amsfonts}
\usepackage{amssymb}
\usepackage{textcomp}
\usepackage{hyphenat}

\biboptions{sort&compress}

\journal{Surface \& Coatings Technology}

\begin{document}

\begin{frontmatter}

\title{Element- and charge-state-resolved ion energies in the cathodic arc plasma from composite AlCr cathodes in argon, nitrogen and oxygen atmospheres}

\author[leoben,berkeley]{Robert Franz\corref{cor}}
\cortext[cor]{Corresponding author.}
\ead{robert.franz@unileoben.ac.at}
\author[plansee]{Peter Polcik}
\author[berkeley]{André Anders}

\address[leoben]{Montanuniversität Leoben, Franz-Josef-Strasse 18, 8700 Leoben, Austria}
\address[plansee]{PLANSEE Composite Materials GmbH, Siebenb\"{u}rgerstrasse 23, 86983 Lechbruck am See, Germany}
\address[berkeley]{Lawrence Berkeley National Laboratory, 1 Cyclotron Road, Berkeley, California 94720, USA}

\date{\today}

\begin{abstract}
The energy distribution functions of ions in the cathodic arc plasma using composite AlCr cathodes were measured as a function of the background gas pressure in the range 0.5 to 3.5 Pa for different cathode compositions and gas atmospheres. The most abundant aluminium ions were Al$^{+}$ regardless of the background gas species, whereas Cr$^{2+}$ ions were dominating in Ar and N$_2$ and Cr$^{+}$ in O$_2$ atmospheres. The energy distributions of the aluminium and chromium ions typically consisted of a high-energy fraction due to acceleration in the expanding plasma plume from the cathode spot and thermalised ions that were subjected to collisions in the plasma cloud. The fraction of the latter increased with increasing background gas pressure. Atomic nitrogen and oxygen ions showed similar energy distributions as the aluminium and chromium ions, whereas the argon and molecular nitrogen and oxygen ions were formed at greater distance from the cathode spot and thus less subject to accelerating gradients. In addition to the positively charged metal and gas ions, negatively charged oxygen and oxygen-containing ions were observed in O$_2$ atmosphere. The obtained results are intended to provide a comprehensive overview of the ion energies and charge states in the arc plasma of AlCr composite cathodes in different gas atmospheres as such plasmas are frequently used to deposit thin films and coatings.
\end{abstract}


\end{frontmatter}

\section{Introduction}

Alloy and composite cathodes are nowadays frequently applied in cathodic arc deposition processes to synthesise various thin film and coating materials. The cathodes supply the non-gasous elements, typically metals, while reactive background gases like N$_2$ and O$_2$ are added during the deposition process to grow nitrides, oxides or mixtures thereof. In order to establish optimal conditions for film growth, a detailed knowledge about the arc plasma properties from such alloy or composite cathodes is of vital importance.

In recent years, several studies have been devoted to measure and analyse the cathodic arc plasma from a wide variety of materials. Chaly \textit{et al.}~reported on the cathode spot motion on different CuCr alloys \cite{Chaly1997}. Models of stationary cathode spots and the near-cathode plasma region on CuCr cathodes with large grains (exceeding the size of the cathode spot) were developed by Almeida \textit{et al.}~\cite{Almeida2013} and Benilov \textit{et al.}~\cite{Benilov2013}. Other cathode spot models based on the explosive nature of the spots, e.g.~the ecton model \cite{Barengolts2003,Mesyats2013}, or on the fractal nature of the spots, e.g.~\cite{Anders2005d}, have not yet been applied to alloy or composite cathodes.

A frequently investigated plasma property is the ion charge state distribution (ICSD). The alloy or composite materials studied include SiC, TiC, TiN, TiO$_2$, WC, brass and stainless steel \cite{Sasaki1989} as well as Ti--80 at.\%Si, Zr--76 at.\%Si and Ni--40 wt.\%Cr--10 wt.\%Al \cite{Eizner1996}. The system Ti--Hf covering the entire compositional range was studied in detail experimentally \cite{Sasaki1993} and theoretically by applying modified Saha equations \cite{Schulke1999}. A relation between the evolution of the ICSD and the electron temperature was pointed out by Savkin \textit{et al.}~when analysing the arc plasma from a series of Cu alloys \cite{Savkin2010}. Pronounced changes in the ICSDs from Bi--Cu cathodes as compared to the single-element cathodes were reported by Adonin \textit{et al.}~\cite{Adonin2012}.

The ion energies in the filtered arc plasma from TiAl cathodes were studied by Bilek \textit{et al.}~\cite{Bilek1998}, where the mean energy strongly decreased with the addition of Al to the cathodes. The ion energy distribution functions (IEDFs) and the plasma composition from TiSi, TiC and TiAl cathodes were analysed by Eriksson \textit{et al.}~~\cite{Eriksson2013} and Zhirkov \textit{et al.}~\cite{Zhirkov2013,Zhirkov2014}. Correlations of the plasma properties with the cohesive energy of the chemical compounds encountered on the cathode surface were noticed. Further, it was shown that for alloy cathodes operated in vacuum arc discharges, all ions regardless of species and charge state had identical most probable velocities. The angular dependence of the ion flow in the arc plasma from alloy and composite cathodes for different background gas pressures was investigated by Nikolaev \textit{et al.}~\cite{Nikolaev2013,Nikolaev2014}.

In a previous work, we reported on the ICSDs in \textit{pulsed} arc plasmas from AlCr cathodes covering the entire compositional range \cite{Franz2013a}. The charge states were measured in dependence on the background gas pressure of Ar, N$_2$ and O$_2$. In the current work, we use dc arc discharges since such plasma conditions are more commonly employed for the synthesis of nitride and oxide thin films and coatings than pulsed arc discharges. The energies of aluminium and chromium ions were measured in dependence on the cathode composition and the background gas pressure. The pressure ranges of N$_2$ and O$_2$ were chosen to be relevant for industrial deposition processes. Argon was added as an inert atmosphere which allows to evaluate possible cathode poisoning effects by comparing the results with the measurements in the reactive atmospheres of N$_2$ and O$_2$. A comprehensive overview of the ion energies and the ion charge states is provided since these are important factors influencing the thin film growth conditions in plasma-based synthesis methods like cathodic arc deposition.

\section{Experimental details}

All experiments were carried out in a vacuum chamber with a diameter of 1 m and a typical residual background pressure of 2.5 $\cdot$ 10$^{-4}$ Pa. A dc arc source (from VTD Vakuumtechnik, Dresden, Germany) with a cathode of a diameter of 65 mm was placed inside the vacuum chamber and operated at an arc current of 50 A during the measurements. The pressure of the gases Ar, N$_2$ or O$_2$ was set prior to plasma ignition to values varying from 0.5 Pa (or 1 Pa in some cases) to 3.5 Pa. In the case of the reactive gases N$_2$ and O$_2$, the pressure dropped by 0.4--0.6 Pa and by 0.5--0.8 Pa, respectively, after the plasma was ignited due to consumption of the gases by the growing film deposited within the chamber. Pressures cited in this work refer to pressure values measured before ignition of arc plasma. The cathodes used for the present investigation comprised single-element Al and Cr cathodes as well as composite cathodes prepared by powder-metallurgical methods with Al/Cr atomic ratios of 75/25, 50/50, and 25/75. The composite and the single-element Al cathodes were manufactured by pressing and forging of Al and Cr powder mixtures at high pressure and high temperature. The Cr cathodes were prepared by hot isostatic pressing at temperatures above 1000 \textcelsius. In all cases, a high material density of at least 99\% of the theoretical density was achieved. With this set of cathodes, the entire compositional range in the Al--Cr system was covered.

After a conditioning time of 5 min at a constant arc current of 50 A, the IEDFs were measured with a differentially pumped, combined ion energy-per-charge and mass-per-charge analyser (\nohyphens{HIDEN} EQP 300). The instrument was set to measure in the energy-per-charge range from 0 to 200 eV/charge with an increment of 0.5 eV/charge for a given mass-to-charge ratio according to the positively charged ion of interest. The dwell time for each data point was 50 ms and the distance between the cathode surface and the orifice of the mass-energy analyser was 24 cm. Subsequent to the measurements, the raw data were corrected for the ion charge state by multiplying the scan voltage with the charge state number and dividing the count rate by the charge state number to account for the energy bin width. No further corrections were made to account for the energy dependence of the acceptance angle and transfer function since they are not well known. The following positively charged ions were analysed: Al$^+$, Al$^{2+}$, Al$^{3+}$, Cr$^+$, Cr$^{2+}$, Cr$^{3+}$, Ar$^+$, Ar$^{2+}$, N$^+$ , N$^{2+}$, N$_2^+$, O$^+$, O$^{2+}$, O$_2^+$, AlO$^+$, CrO$^+$. In addition, the negatively charged ions O$^-$, O$_2^-$, AlO$^-$ and CrO$^-$ were measured in the energy-per-charge range from 0 to 150 eV/charge with an increment of 0.5 eV/charge. In this case, a multiplier voltage of 2600 V was used, as compared to 1675 V in the case of the positively charged ions, in order to enhance the intensity of the recorded negative ion signal.

\section{Results}

\subsection{Argon atmosphere}

\begin{figure*}[!t]
 \centering
 \includegraphics[width=16cm]{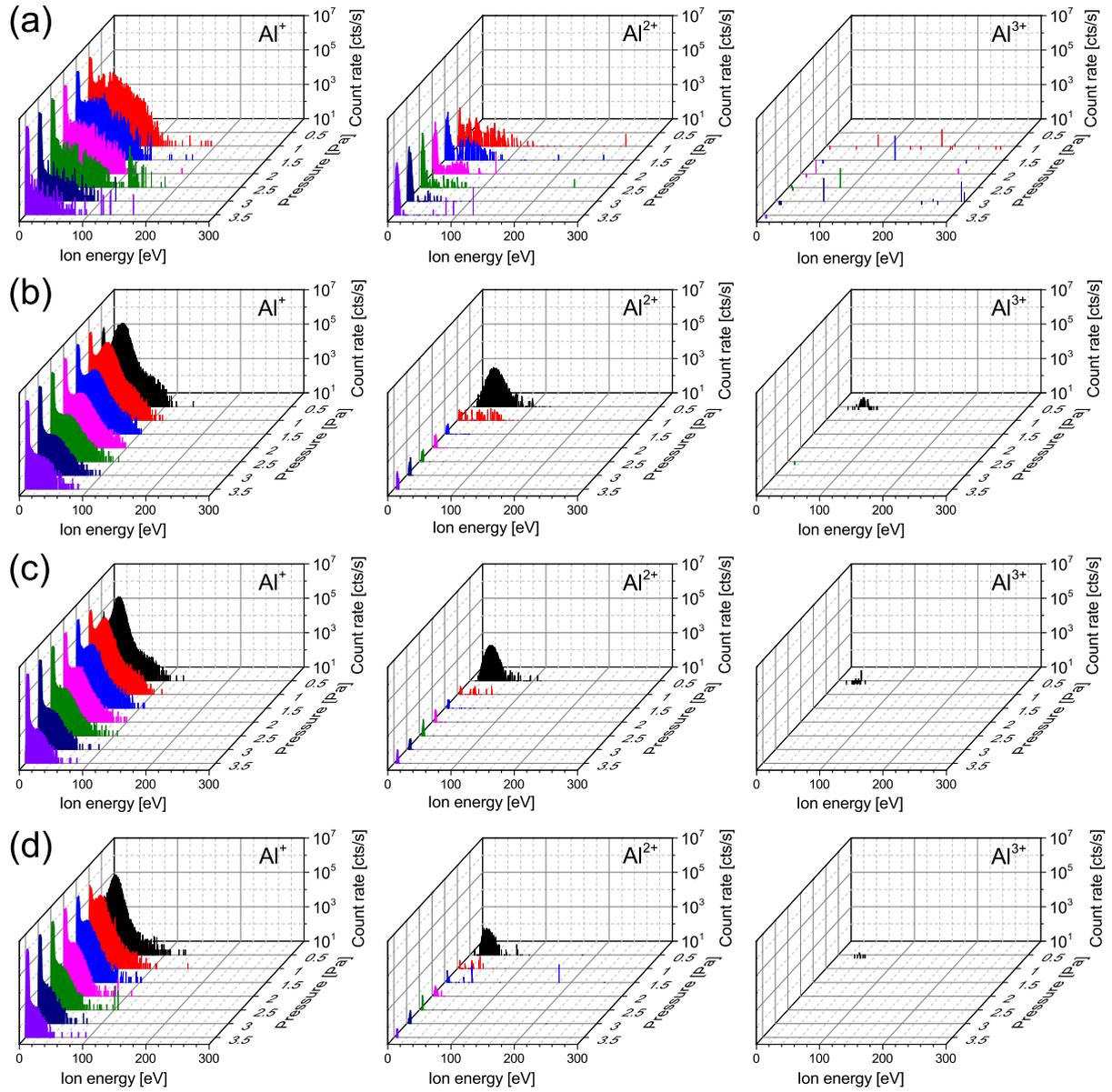}
 \caption{IEDFs of aluminium ions in the arc plasma of (a) single-element Al, (b) AlCr(75/25), (c) AlCr(50/50) and (d) AlCr(25/75) cathode in Ar atmosphere.}
 \label{fig:Ar_Al-ions_IEDFs}
\end{figure*}

As shown in Fig.~\ref{fig:Ar_Al-ions_IEDFs}, the dominating aluminium ion species in Ar atmosphere is Al$^+$ regardless of the cathode composition or Ar pressure. Small fractions of Al$^{2+}$ ions were observed in the plasma from the single-element Al cathode or at a low Ar pressure of 0.5 Pa. The discharge from the single-element Al cathode at only 0.5 Pa was not stable and is, hence, not shown here. All Al$^+$ IEDFs consist of thermalised ions and a high-energy part that extends up to 100 eV. As can be expected, the fraction of the thermalised ions increases with increasing Ar pressure, whereas the energy of the peak in the energetic portion of the IEDF decreases with decreasing Al content.

\begin{figure*}[!t]
 \centering
 \includegraphics[width=16cm]{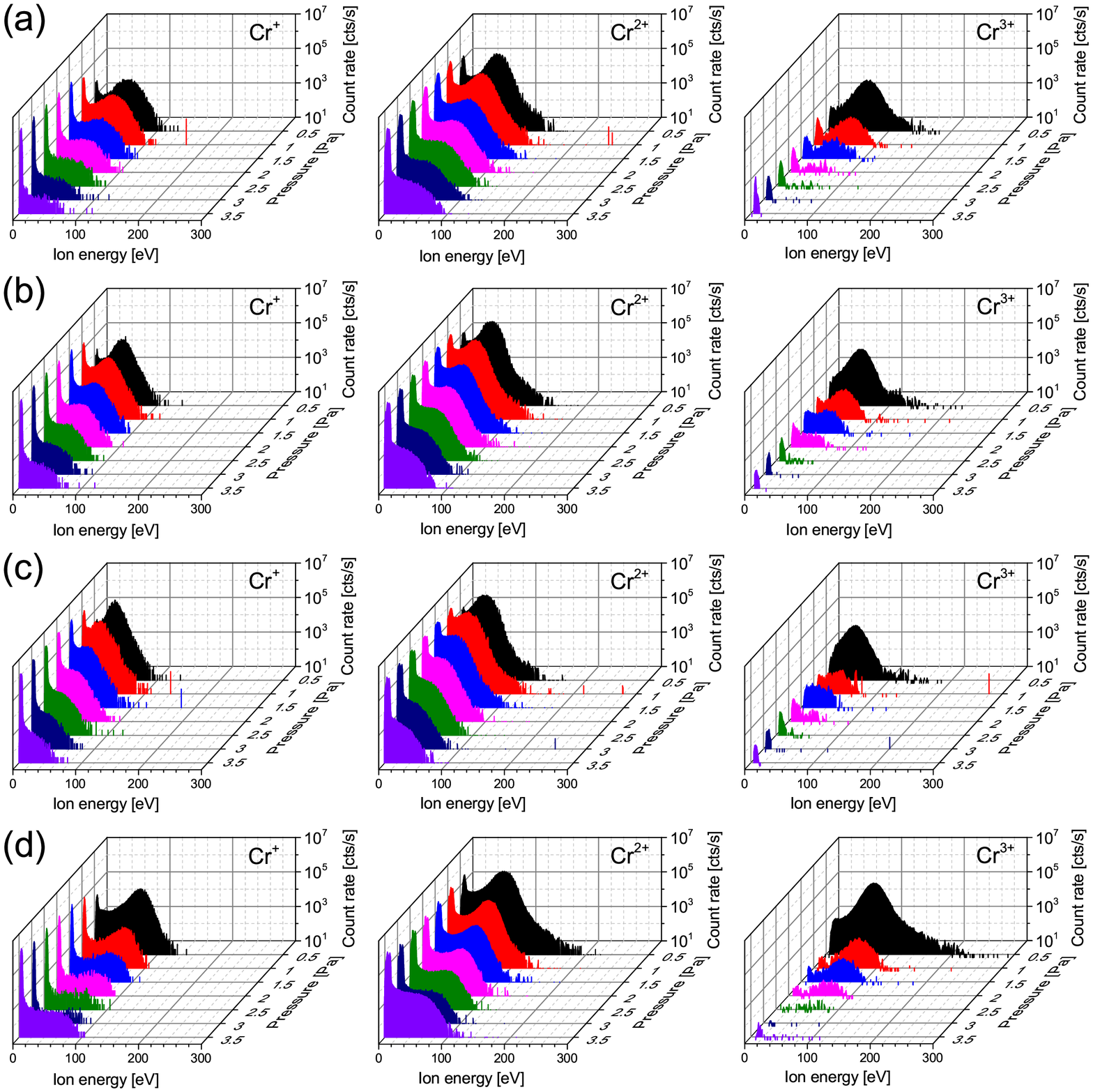}
 \caption{IEDFs of chromium ions in the arc plasma of (a) AlCr(75/25), (b) AlCr(50/50), (c) AlCr(25/75) and (d) single-element Cr cathode in Ar atmosphere.}
 \label{fig:Ar_Cr-ions_IEDFs}
\end{figure*}

In the case of chromium ions, the IEDFs shown in Fig.~\ref{fig:Ar_Cr-ions_IEDFs} are similar. However, the dominating chromium ion species is Cr$^{2+}$ in all discharge conditions studied. Further, the maximum energy of the high-energy tail reaches higher values than in the case of aluminium and extends up to 150 eV (in some cases even up to 200 eV). The energy of the peak in the non-thermal part of the distribution is highest for the single-element Cr cathode and increases with decreasing Cr content in the composite cathodes. The influence of the Ar background gas is again noticed by in an increase in the fraction of thermalised ions and a decrease in the fraction of Cr$^{3+}$ ions as the gas pressure increases.

\begin{figure*}[!t]
 \centering
 \includegraphics[width=11cm]{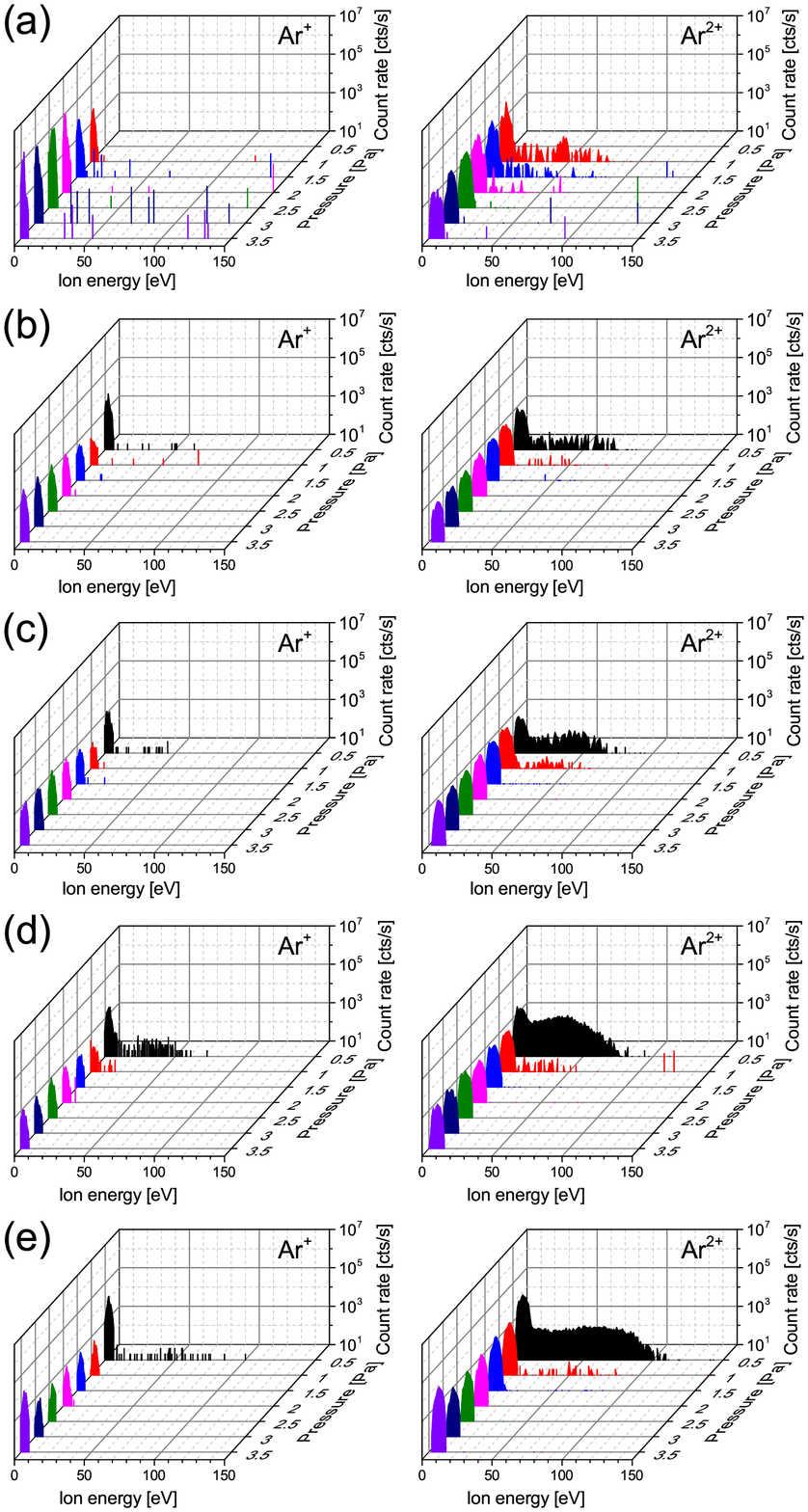}
 \caption{IEDFs of argon ions in the arc plasma of (a) single-element Al, (b) AlCr(75/25), (c) AlCr(50/50), (d) AlCr(25/75) and (e) single-element Cr cathode in Ar atmosphere.}
 \label{fig:Ar_Ar-ions_IEDFs}
\end{figure*}

All Ar$^+$ and Ar$^{2+}$ ions (see Fig.~\ref{fig:Ar_Ar-ions_IEDFs}) were all thermalised with the exception when Cr-rich cathodes were used with a low Ar pressure of 0.5 Pa. In this case, a high-energy tail for the Ar$^{2+}$ ions was observed with the maximum energy reaching up to 100 eV.

\subsection{Nitrogen atmosphere}

\begin{figure*}[!t]
 \centering
 \includegraphics[width=16cm]{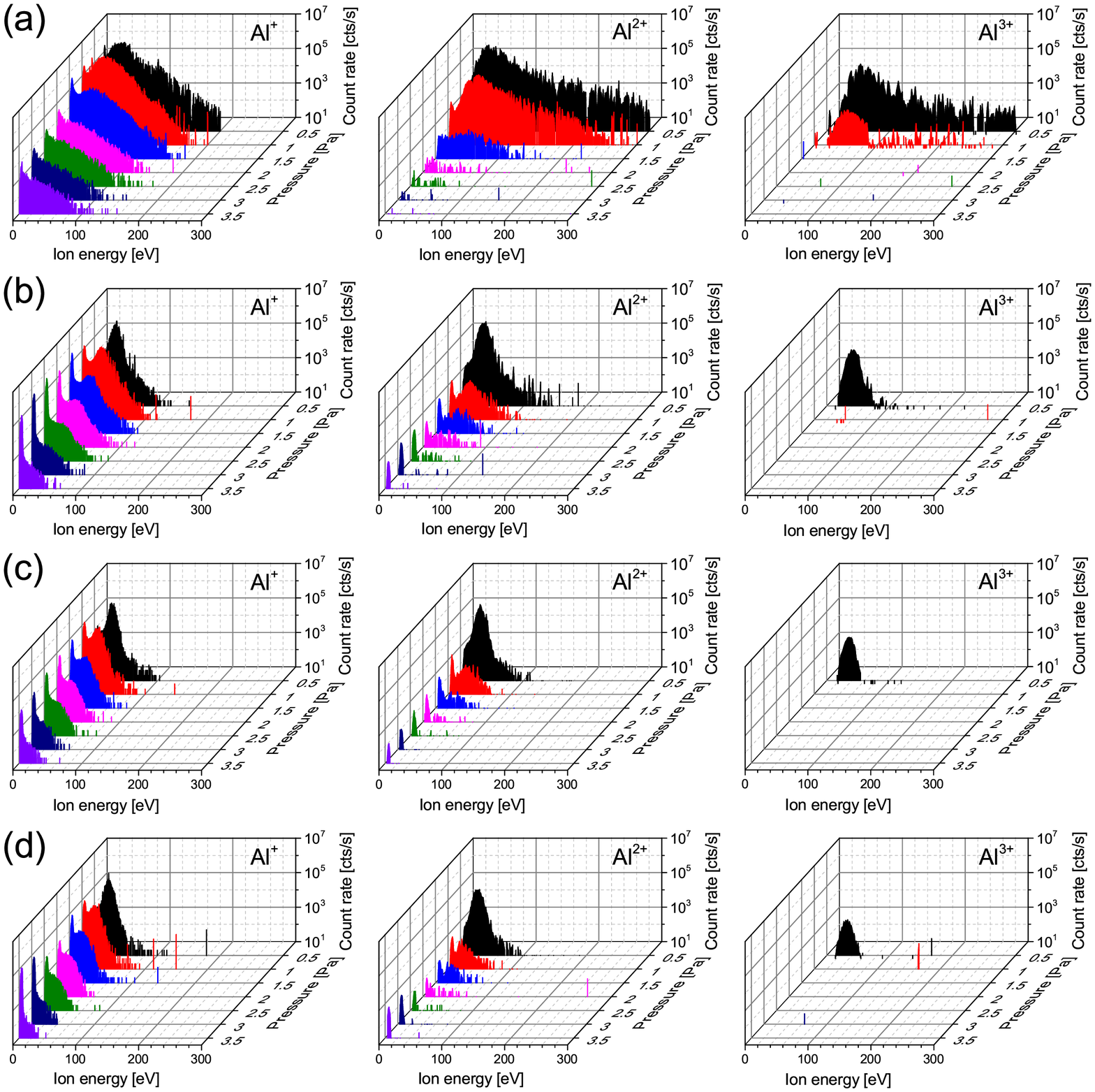}
 \caption{IEDFs of aluminium ions in the arc plasma of (a) single-element Al, (b) AlCr(75/25), (c) AlCr(50/50) and (d) AlCr(25/75) cathode in N$_{2}$ atmosphere.}
 \label{fig:N2_Al-ions_IEDFs}
\end{figure*}

\begin{figure*}[!t]
 \centering
 \includegraphics[width=16cm]{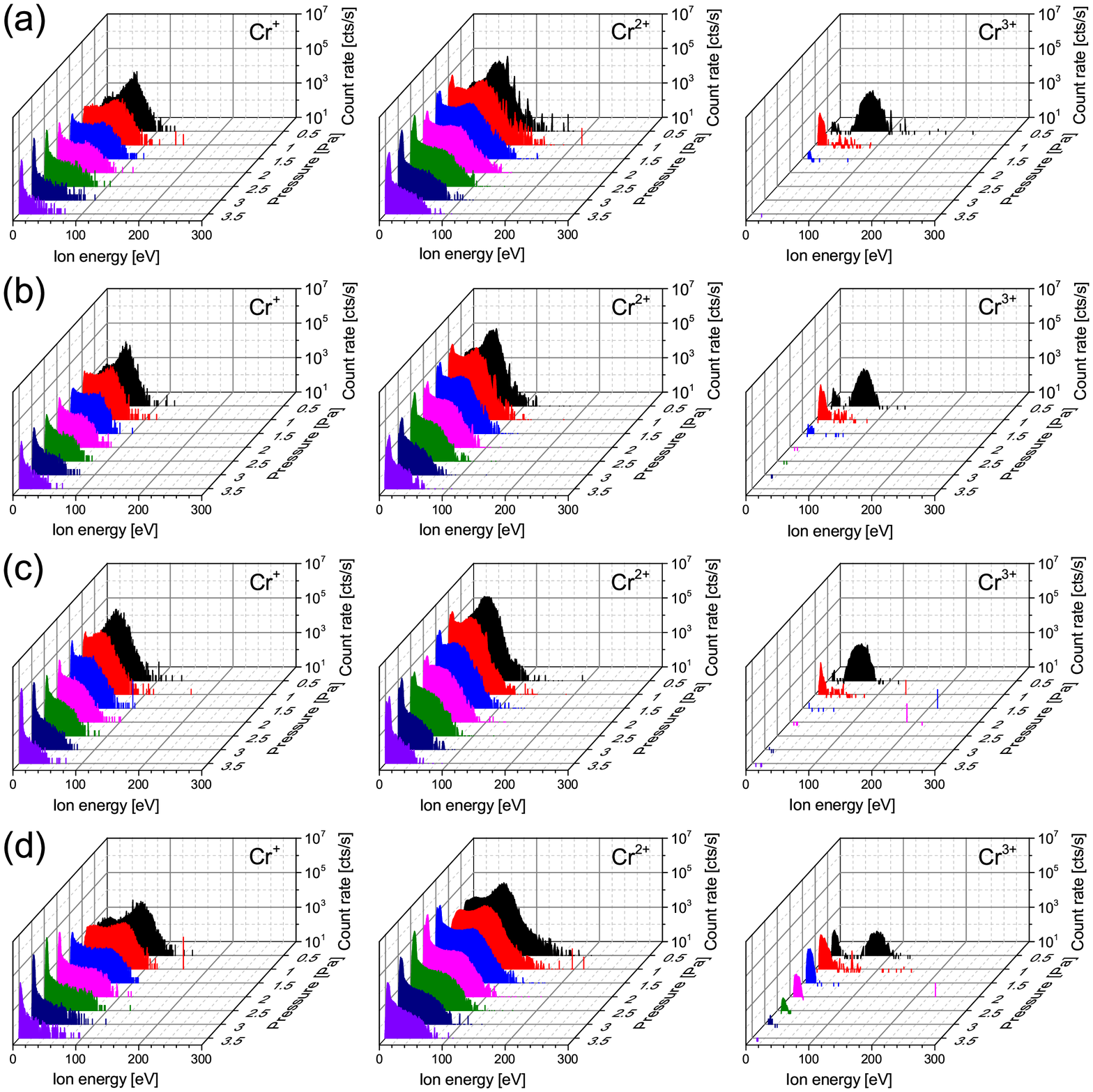}
 \caption{IEDFs of chromium ions in the arc plasma of (a) AlCr(75/25), (b) AlCr(50/50), (c) AlCr(25/75) and (d) single-element Cr cathode in N$_{2}$ atmosphere.}
 \label{fig:N2_Cr-ions_IEDFs}
\end{figure*}

The IEDFs of the aluminium and chromium ions in N$_2$ atmosphere (see Figs.~\ref{fig:N2_Al-ions_IEDFs} and \ref{fig:N2_Cr-ions_IEDFs}) are generally comparable to the IEDFs in Ar atmosphere. The differences include a higher fraction of multiply charged aluminium ions at low gas pressures and more pronounced high-energy tails that extend up to 200--300 eV. The aluminium IEDFs in the plasma from the composite cathode show a higher fraction of thermalised ions, but the reduction in the charge states with increasing N$_2$ pressure is similar for all cathode compositions. In the case of chromium, less Cr$^{3+}$ were recorded in N$_2$ than in Ar atmosphere, but the Cr$^{2+}$ ions are still the most abundant chromium ions.

\begin{figure*}[!t]
 \centering
 \includegraphics[width=11cm]{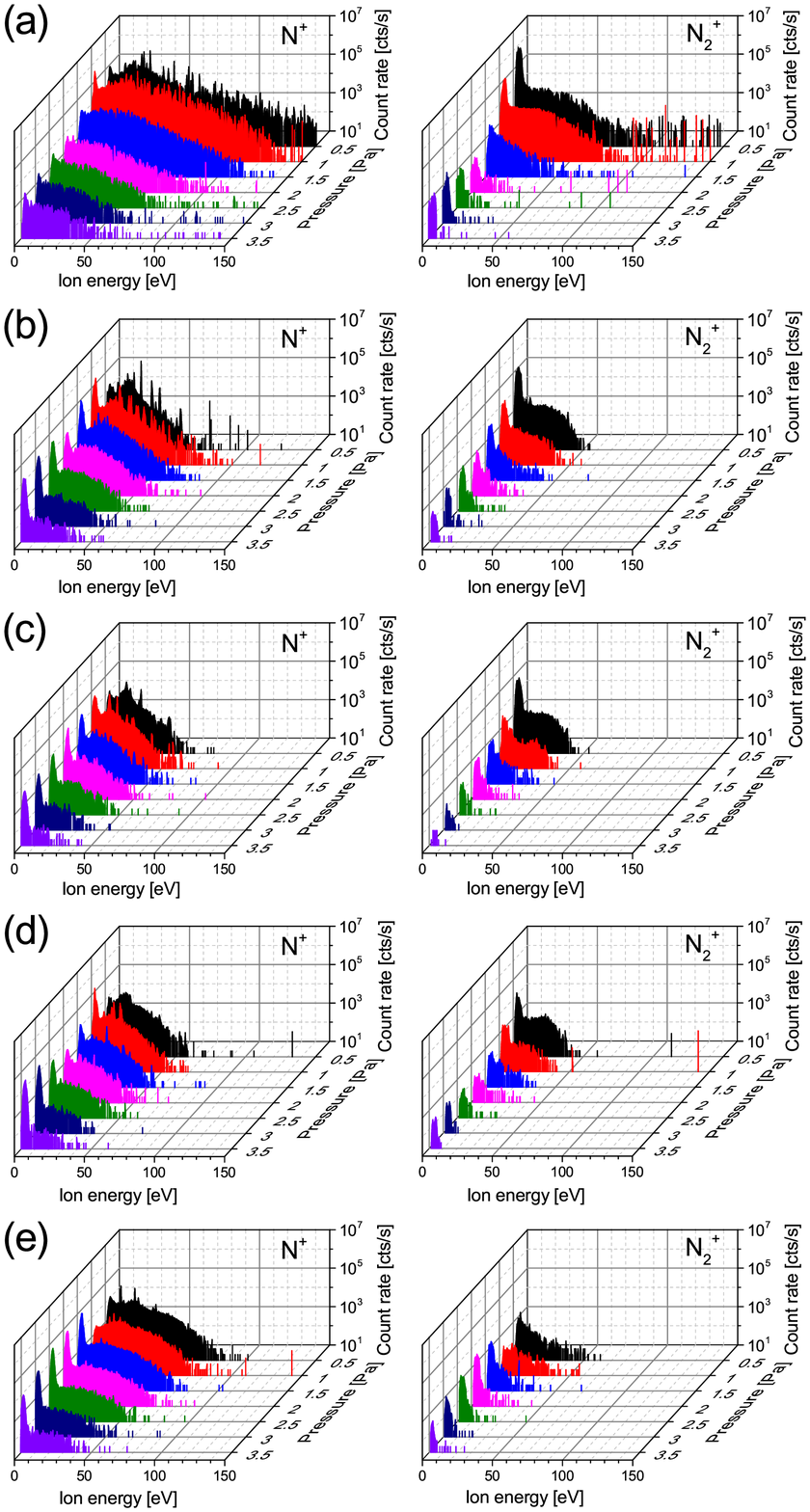}
 \caption{IEDFs of nitrogen ions in the arc plasma of (a) single-element Al, (b) AlCr(75/25), (c) AlCr(50/50), (d) AlCr(25/75) and (e) single-element Cr cathode in N$_{2}$ atmosphere.}
 \label{fig:N2_N-ions_IEDFs}
\end{figure*}

A major difference was observed between the IEDFs of the gas ions in Ar and N$_2$ atmospheres. Three different nitrogen ions were encountered: N$^+$ and N$_2^+$ (see Fig.~\ref{fig:N2_N-ions_IEDFs}) as well as small fractions of N$^{2+}$ (not shown). In contrast to the argon ions, the nitrogen ions show pronounced high-energy tails, especially in the case of N$^+$ where the maximum energy can reach values up to 100 eV. 

\subsection{Oxygen atmosphere}

\begin{figure*}[!t]
 \centering
 \includegraphics[width=16cm]{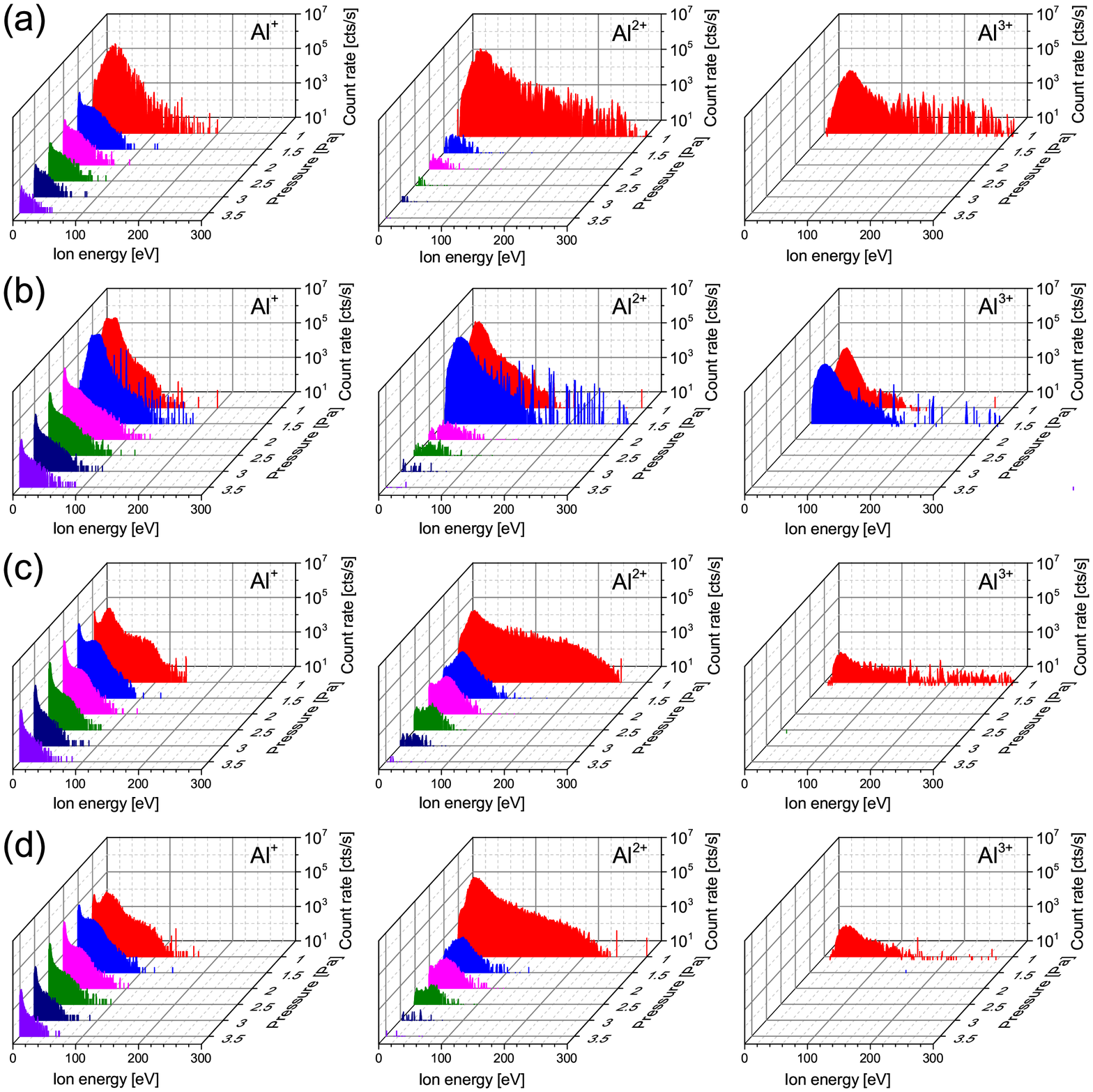}
 \caption{IEDFs of aluminium ions in the arc plasma of (a) single-element Al, (b) AlCr(75/25), (c) AlCr(50/50) and (d) AlCr(25/75) cathode in O$_{2}$ atmosphere.}
 \label{fig:O2_Al-ions_IEDFs}
\end{figure*}

While there are pronounced contributions in the IEDFs of all aluminium ions at low O$_2$ pressure (see Fig.~\ref{fig:O2_Al-ions_IEDFs}), the number of the multiply charged aluminium ions is strongly reduced upon increasing the pressure. However, a considerable fraction of Al$^{2+}$ ions is still present in the plasma from Cr-rich cathodes at medium O$_2$ pressures, which is in contrast to the other two atmospheres studied. Apart from theses differences, the aluminium IEDFs in O$_2$ atmosphere are similar to Ar and N$_2$ atmosphere.

\begin{figure*}[!t]
 \centering
 \includegraphics[width=16cm]{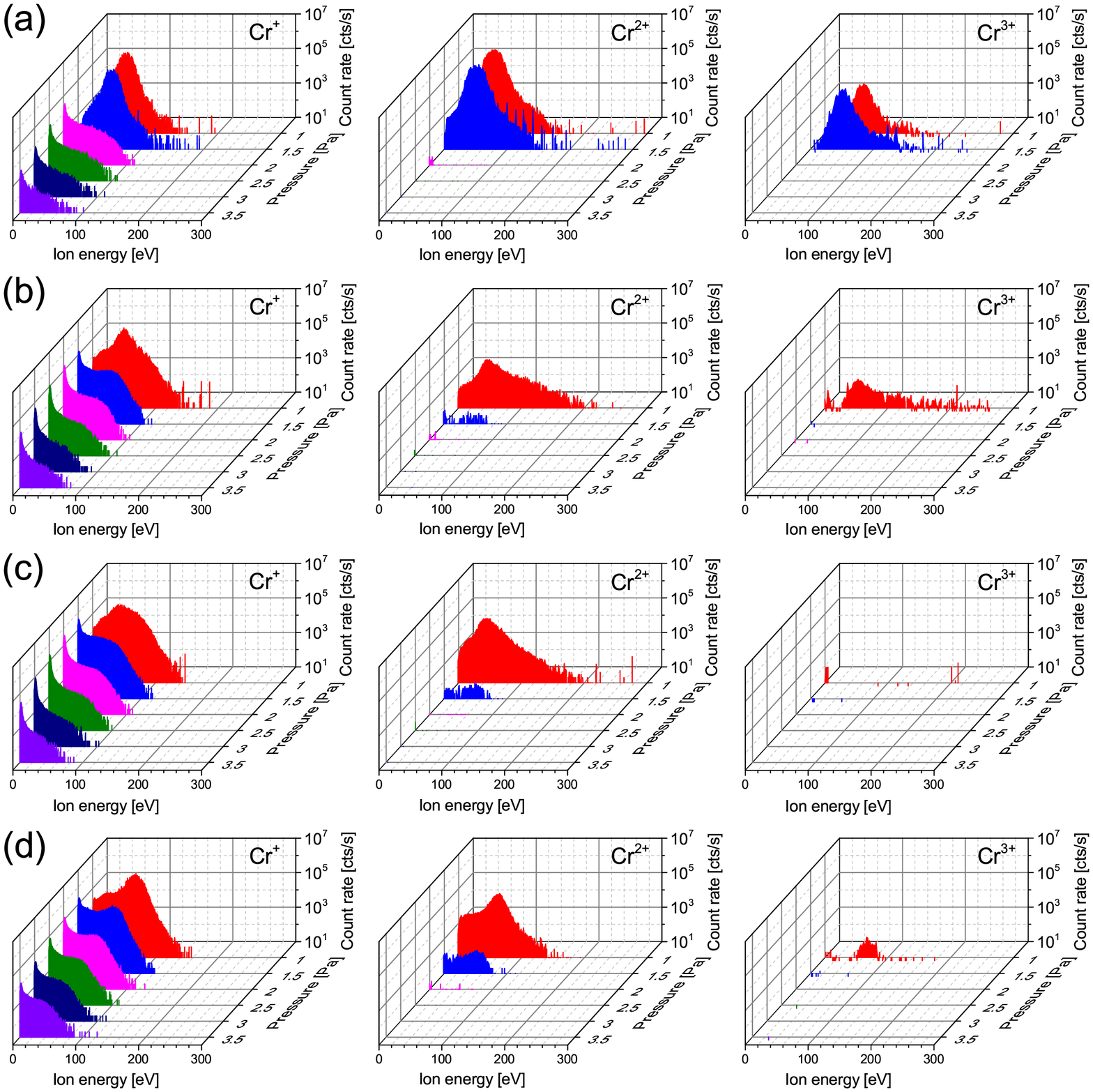}
 \caption{IEDFs of chromium ions in the arc plasma of (a) AlCr(75/25), (b) AlCr(50/50), (c) AlCr(25/75) and (d) single-element Cr cathode in O$_{2}$ atmosphere.}
 \label{fig:O2_Cr-ions_IEDFs}
\end{figure*}

As shown in Fig.~\ref{fig:O2_Cr-ions_IEDFs}, the most abundant chromium ion in O$_2$ is Cr$^{+}$ for all studied cathodes. Multiply charged chromium ions were only observed at O$_2$ pressures below 2 Pa.

\begin{figure*}[!t]
 \centering
 \includegraphics[width=11cm]{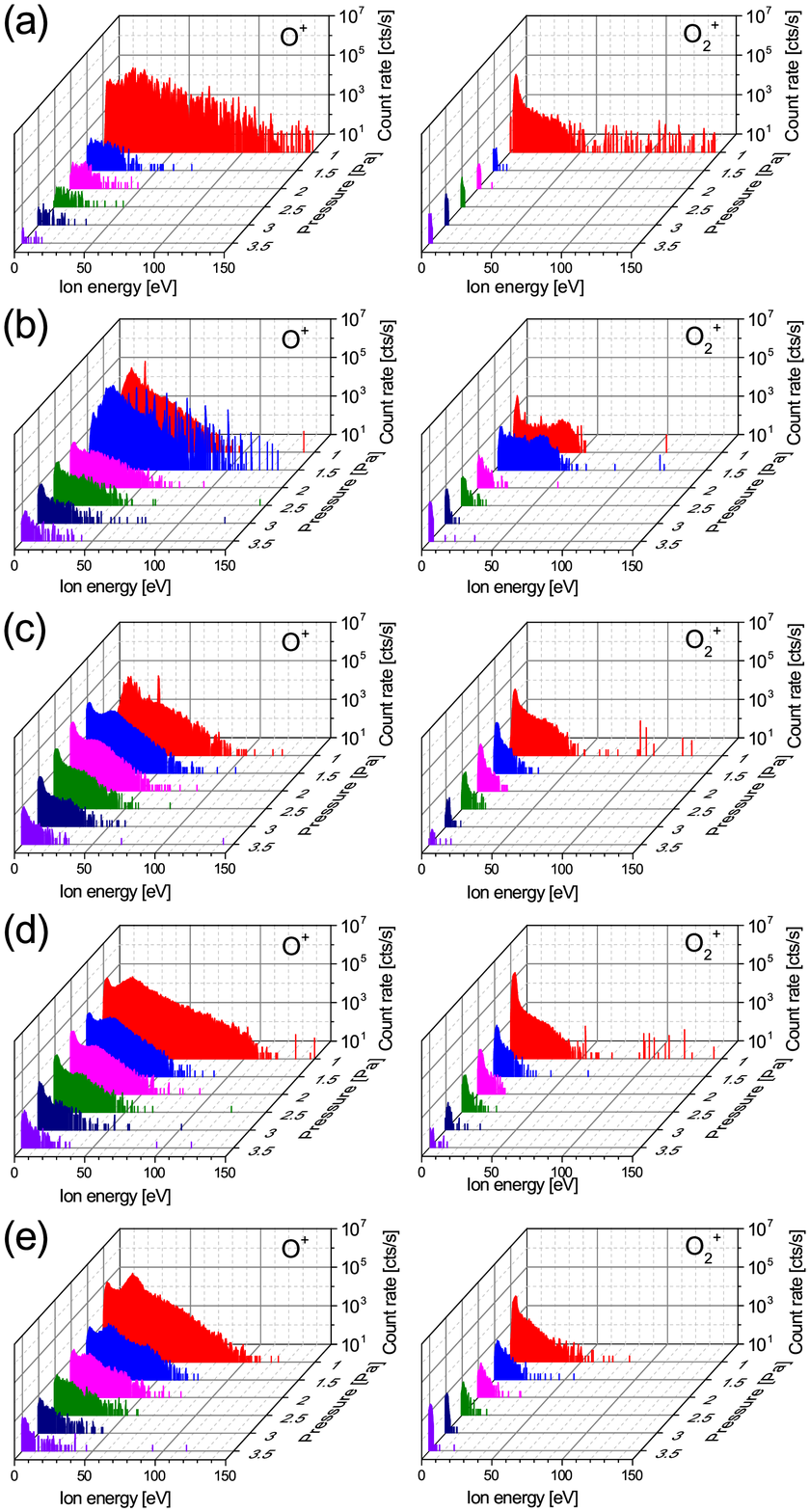}
 \caption{IEDFs of positively charged oxygen ions in the arc plasma of (a) single-element Al, (b) AlCr(75/25), (c) AlCr(50/50), (d) AlCr(25/75) and (e) single-element Cr cathode in O$_{2}$ atmosphere.}
 \label{fig:O2_positive-O-ions_IEDFs}
\end{figure*}

\begin{figure*}[!t]
 \centering
 \includegraphics[width=11cm]{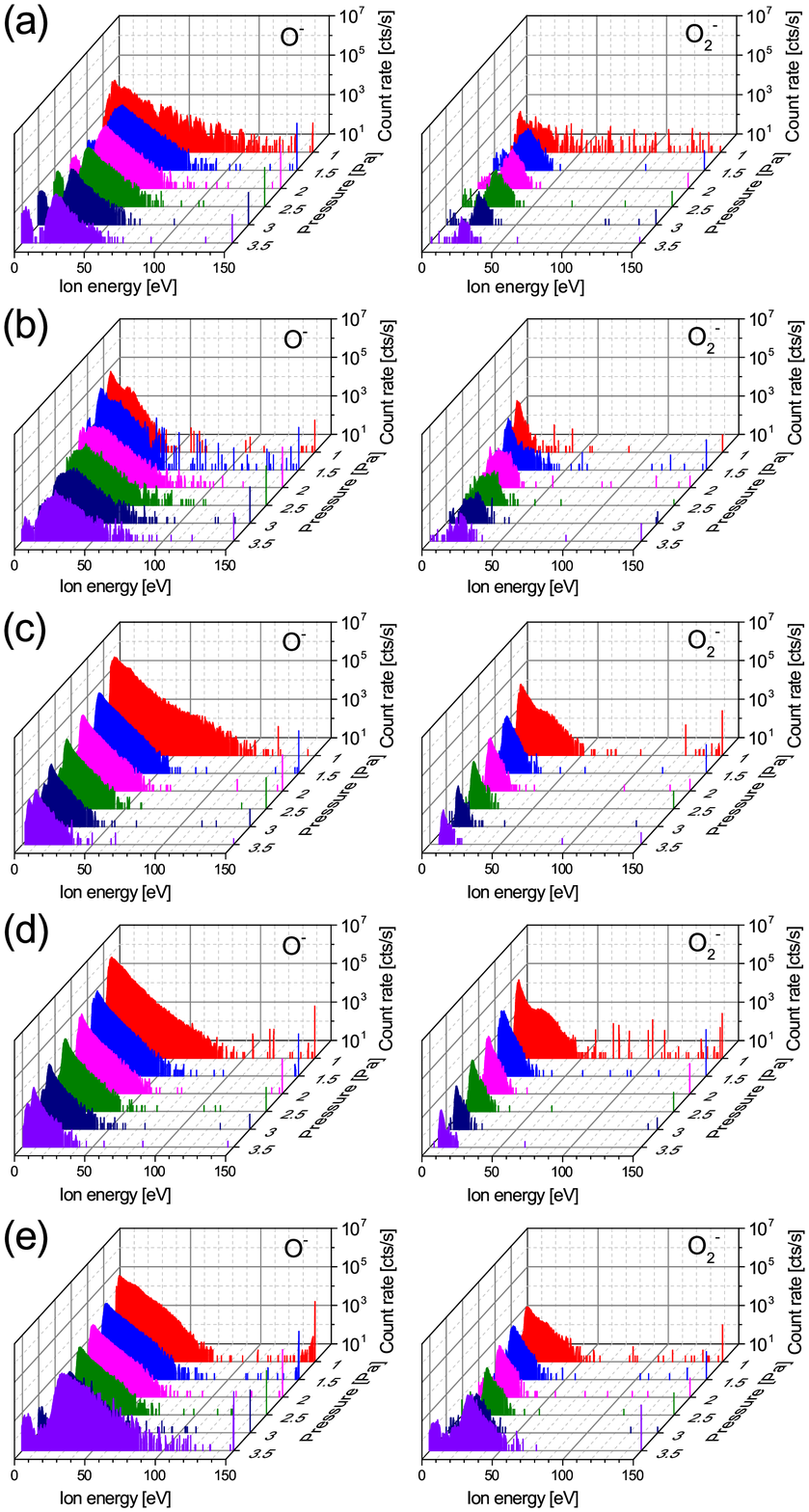}
 \caption{IEDFs of negatively charged oxygen ions in the arc plasma of (a) single-element Al, (b) AlCr(75/25), (c) AlCr(50/50), (d) AlCr(25/75) and (e) single-element Cr cathode in O$_{2}$ atmosphere.}
 \label{fig:O2_negative-O-ions_IEDFs}
\end{figure*}

A major difference of the plasma properties in O$_2$ atmosphere as compared to Ar and N$_2$ is a series of oxygen-containing molecular ions. Figs.~\ref{fig:O2_positive-O-ions_IEDFs} and \ref{fig:O2_negative-O-ions_IEDFs} show atomic and molecular oxygen ions that are positively and negatively charged, respectively. While the O$_2^+$ ions are mainly thermalised, the O$^+$ ions show a pronounced high-energy tail and their number increases with increasing Cr content in the cathode. At an O$_2$ pressure of 1 Pa a small fraction of O$^{2+}$ ions was also observed (not shown). Comparing the negatively charged atomic and molecular oxygen ions (see Fig.~\ref{fig:O2_negative-O-ions_IEDFs}), a similar behaviour as in the case of the positively charged ions can be noticed. The maximum energy in the IEDFs of the O$^-$ ions reaches values of up to 50 eV, in few cases even 100 eV. However, as an estimate based on the integrated ion counts (recorded at a constant multiplier voltage) over all energies, only about 1\% of the oxygen-containing ions in the plasma are negatively charged.

\begin{figure*}[!t]
 \centering
 \includegraphics[width=11cm]{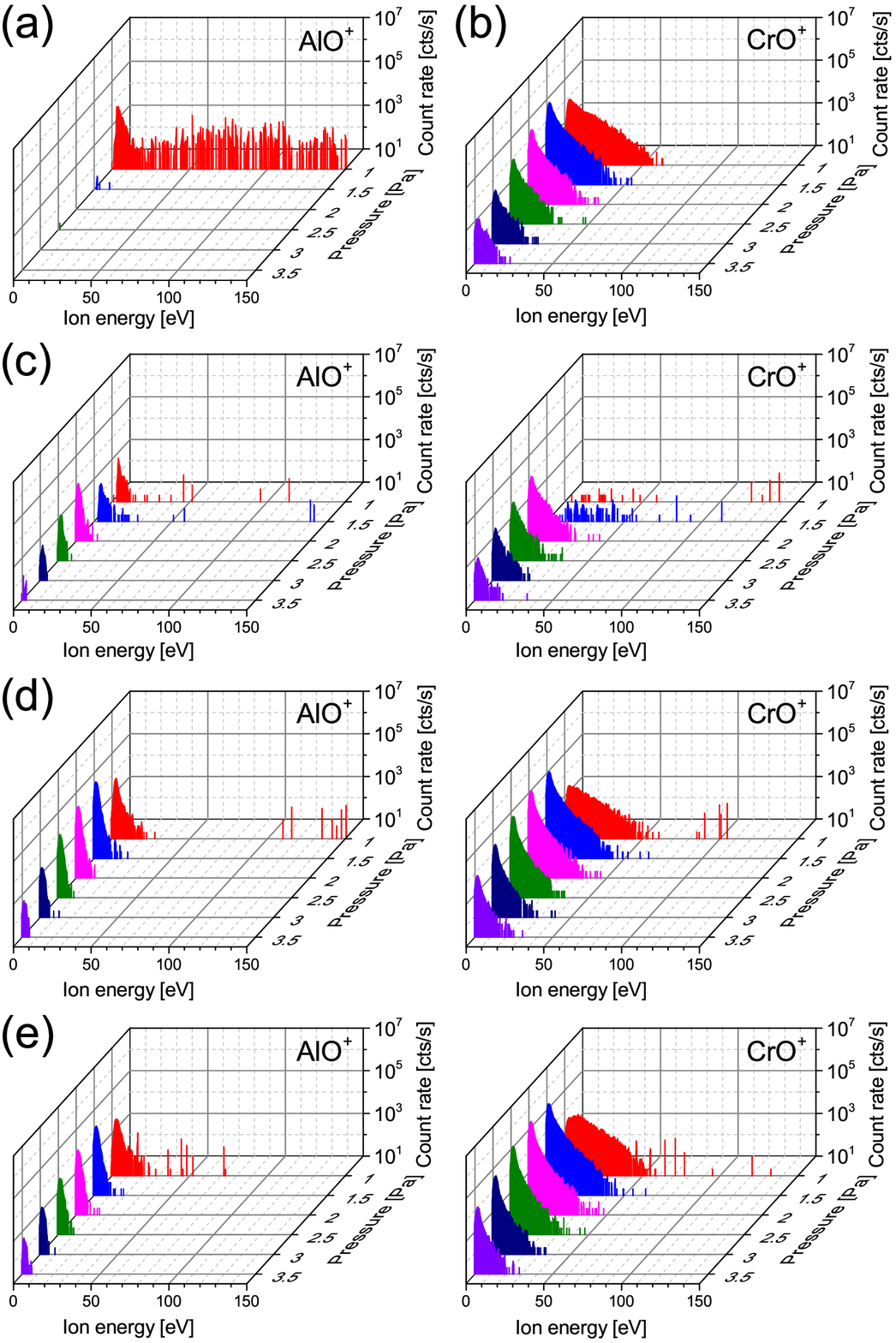}
 \caption{IEDFs of positively charged metal oxide ions in the arc plasma of (a) single-element Al, (b) single-element Cr cathode, (c) AlCr(75/25), (d) AlCr(50/50) and (e) AlCr(25/75) cathode in O$_{2}$ atmosphere.}
 \label{fig:O2_positive-MeO-ions_IEDFs}
\end{figure*}

An interesting feature of the oxygen-containing plasma from the AlCr cathodes are the metal oxide ions as shown in Fig.~\ref{fig:O2_positive-MeO-ions_IEDFs}. The AlO$^+$ ions are mainly formed in the plasma from the composite cathodes and their energy distribution shows that they are thermalised. In contrast, CrO$^+$ are present in a higher number and the maximum energy in their IEDFs can reach up to 50 eV. At an O$_{2}$ pressure of 1 Pa traces of AlO$^-$ ions were also noticed (not shown), whereas CrO$^-$ ions were absent in the entire O$_{2}$ pressure range studied.

\section{Discussion}

As mentioned above, the intention of the current work is to provide a comprehensive overview of the arc plasma properties from AlCr cathodes in different atmospheres. However, emphasis in the discussion of the results is given to general trends over specific details or effects observed in subsets of the data.

\subsection{Evolution of ion charge states}

The effects observed regarding the ion charge states are similar to our previous work on the pulsed arc plasma from AlCr cathodes \cite{Franz2013a}. A detailed discussion about factors influencing the evolution of the ion charge states can be found there. Here, a short summary will be given, mainly pointing out differences in the results from the discharge conditions used.

In general, the ICSDs are altered due to a convolution of effects associated with the background gas and the cathode composition. The influence of the cathode composition is less pronounced and better visible at low gas pressures close to high vacuum. The aluminium charge states are basically independent of the cathode composition, whereas the chromium charge states are reduced when Al is added to the cathode. However, with increasing gas pressure the fraction of all multiply charged ions is reduced due to charge exchange collisions \cite{Franz2013a}. This reduction is similar in Ar and N$_2$ atmospheres indicating that poisoning effects are without strong influence on the charge states in N$_2$ atmosphere. In O$_2$ atmosphere, the aluminium charge states are similar to the other atmospheres studied, but the multiply charged chromium ions are almost completely lost. This points towards a pronounced, yet selective influence of cathode poisoning on the ion charge states in O$_2$ atmosphere.

A major difference between our previous \cite{Franz2013a} and the current work is due to the ionisation of the background gas. In the employed dc arc source, a magnetic field is used to steer the cathode spot along a circular erosion track. Steering of the cathode spot is typically applied to accelerate the cathode spot motion which in turn results in a reduced emission of macroparticles that can cause detrimental effects if incorporated into the growing film \cite{Anders2008}. However, in the current case, the presence of a magnetic field causes a significant ionisation of the background gas as reported by Oks \textit{et al.}~\cite{Oks1995,Oks1996b}. In all atmospheres used within this work, mainly single-charged atomic and molecular gas ions were observed. In the case of O$_2$ atmosphere, single-charged AlO$^+$ and CrO$^+$ ions were noticed in addition as well as negatively charged oxygen and oxygen-containing ions.

\subsection{Evolution of aluminium and chromium ion energies}

In general, the energy gain of the ions in cathodic arc plasmas is due to the adiabatic expansion of the plasma plume from the \textmu m-sized cathode spot into the vacuum chamber \cite{Anders2008}. This results in an ion velocity almost independent of the charge state for a given cathode material \cite{Yushkov2000a}. In the case of alloy or composite cathodes, the ion velocity is also independent of the element as pointed out by Zhirkov \textit{et al.}~\cite{Zhirkov2013}. Due to the different mass of the elements, the (kinetic) energy distributions differ from one another. This so-called 'velocity rule' appears also to be valid for the here-studied AlCr cathodes at low pressures where a distinct peak (with higher energy than thermalised ions) in the IEDFs can be determined and the most probable velocity calculated. However, the absolute value of this velocity is affected by processes within the cathode spot, i.e.~at the interface between cathode surface and plasma region. The so-called \textit{cohesive energy rule} relating the cathode material properties to the plasma properties has been established for single-element cathodes \cite{Anders2001a}. In the case of alloy or composite cathodes, the frequent heating and cooling of the near-surface region due to the travelling cathode spot typically results in the formation of a series of compounds, in the present case intermetallics, nitrides and oxides, which complicates a direct relation between the cathode material and plasma properties.
 
The abovementioned effects are valid for vacuum or low pressure conditions with mean free paths exceeding the vacuum system dimensions. However, the most apparent result regarding the ion energies is the reduction of high-energy ions when the gas pressure is increased. The increased probability for collisions at higher gas pressures results in a slow-down of the ions and, hence, in an increase in the number of thermalised ions as it was described, e.g., for Ti$^+$ ions in a filtered cathodic arc plasma in \cite{Bilek1998}. This effect appears to be independent of the charge state as, e.g., the IEDFs of both Cr$^{+}$ and Cr$^{2+}$ in Ar as well as N$_2$ atmosphere show pronounced contributions from thermalised ions at pressures exceeding 2 Pa. In addition to the reduction in energy, the absolute number of aluminium and chromium ions also decreases when the gas pressure is increased. This affects the film growth conditions especially if a bias voltage is applied to the substrate, which is common practice in the deposition of thin films and coatings using cathodic arcs. A reduced number of available ions at higher pressure results in a reduced ion bombardment of the growing film and a reduced charge state is also important because the ion charge state number is a multiplier between applied substrate voltage and resulting energy.

\subsection{Gas atomic and molecular ions}

When analysing the IEDFs of the gas atomic and molecular ions, their shape can be interpreted with respect of the origin of the ions, in particular the ions with high energy. There are two different processes imaginable how these ions can gain high energy. On the one hand, they can get accelerated during the expansion of the plasma plume from the cathode spot, i.e.~the energy gain is equal to the metal ions from the cathode material. On the other hand, gas atomic and molecular ions can gain energy by momentum transfer in collisions with fast metal ions. Both processes will be discussed briefly in the following taking into account the results obtained.

At low N$_2$ and O$_2$ pressure, the IEDFs of N$^+$ and O$^+$, respectively, have a shape similar to the IEDFs of the respective aluminium and chromium ions. The peak in the energy distributions between 15 and 30 eV corresponds to the abovementioned most probable velocity for these ions. This indicates that the ions were formed in the cathode spot and then accelerated as the plasma cloud expanded. Most likely, the ions originate from atoms or molecules that were attached to the cathode surface prior to the ignition of the plasma in the cathode spot and then were ionised. Gas atoms and molecules can attach to the cathode surface by physisorption or chemisorption. The latter case, where chemical bonds are formed between the gas particles and the cathode material, is known as cathode poisoning.

Another contribution to the IEDFs is due to thermalised ions having energies below 10 eV. Such ions are formed in the already extended plasma cloud or were slowed down after a series of collisions with other ions or atoms. The IEDFs of the measured Ar$^{+}$ and Ar$^{2+}$ as well as N$_2^+$ and O$_2^+$ ions mainly show thermalised ions even at low gas pressures indicating that these ions were formed after the plasma plume expanded from the cathode spot. In the case of argon, a noble gas, atoms do not adsorb to the surface of the cathodes in great quantity. Therefore, ionisation of argon atoms mainly occurs in the expanded plasma cloud where they are not subject to acceleration mechanisms associated with the strong pressure gradients of the cathode spot. In the case of molecular N$_2^+$ and O$_2^+$ ions, one should keep in mind that a layer of oxide or nitride, depending on the gas present, will be eroded by the action of the cathode spot. The high power density in and near the cathode spot leads to removal of material from the cathode in the form of atoms and, most likely, atomic ions, not molecules. Should molecules be present they would be dissociated. At distances large compared to the size of the spot, molecules could form but they are not in the region of strong gradients, hence their acceleration is small. This explains the lack of a high-energy fraction of these molecular ions.

Mainly low-energy, thermalised ions were also recorded in the case of AlO$^+$ ions, whereas the IEDFs of the CrO$^+$ ions show a notable high-energy tail indicating that some CrO$^+$ ions are formed relatively close to the cathode spot. This observation is in agreement with the abovementioned effect that cathode poisoning in O$_2$ atmosphere mainly affects the chromium ions. It seems that oxygen atoms preferentially attach to chromium in the composite cathodes and, hence, the ions get accelerated in the cathode spot plasma expansion the same way metal ions get accelerated.

\subsection{Negatively charged oxygen ions}

Similar considerations as for the gas atomic and molecular ions can be done for the negatively charged oxygen ions. In the cathodic arc plasma the cathode fall potential drop is up to a few 10 V and, therefore, generally causes only a weak acceleration of negatively charged particles. This situation is in contrast to magnetron sputtering plasmas where negatively charged oxygen ions of a few 100 eV can be observed due to the high potential drop in the cathode fall (see e.g.~\cite{Mraz2006a}). While the IEDFs of O$^-_2$ show mainly contributions from thermalised ions, the IEDFs of O$^-$ extend to 50 eV (in some cases even up to 100 eV). The appearance of such high ion energies can be understood by the fact that these ions were also accelerated in the plasma plume expansion, even though acceleration of negatively charged ions during fluctuations of the cathode fall potential drop, i.e.~when the voltage is transiently higher than the average value, cannot be excluded completely. However, the overall number of negatively charged oxygen ions is small with only about 1\% of the entire oxygen-containing ions observed and the effect on film growth is therefore expected to be insignificant.

\section{Conclusions}

From the comprehensive overview of the ion energies in the cathodic arc plasma of composite AlCr cathodes several conclusions can be drawn. As expected, and in agreement with the vast literature on arc plasmas, multiply charged ions were encountered. However, Al$^+$ was the most abundant aluminium ion in all studied atmospheres from 0.5 to 3.5 Pa of Ar, N$_2$, and O$_2$, whereas Cr$^{2+}$ was the dominanting chromium ion in Ar and N$_2$. The fact that Cr$^+$ was the most abundant chromium ion in O$_2$ atmosphere is most likely related to cathode poisoning effects, i.e.~the formation of oxide phases on the cathode surface.

In terms of ion energy, high-energy tails were recorded in the ion energy distribution functions of aluminium and chromium ions as well as N$^+$ and O$^+$ ions. High ion energies originate from acceleration of the cathode spot plasma during expansion into the vacuum chamber. With increasing background gas pressure more collisions occur in the plasma cloud resulting in a slow-down of the ions and an increase of the fraction of thermalised ions having an energy of a few eV. This process occurred for all gas species used.

In O$_2$ atmosphere, negatively charged oxygen and oxygen-containing ions were observed in addition to the otherwise positively charged metal and gas ions. These ions make up a fraction of about 1\% of all oxygen and oxygen-containing ions. Their energy distributions typically extend up to 50 eV indicating an acceleration of these ions in the expansion of the plasma plume with a possible contribution of acceleration by the cathode fall voltage, which is known to be low but fluctuating.

\section*{Acknowledgements}

R.~Franz gratefully acknowledges the support of an Erwin Schrödinger Fellowship by the Austrian Science Fund (FWF, Project J3168-N20) which enabled his research at LBNL. Work at LBNL is supported by the U.S.~Department of Energy under Contract No.~DE-AC02-05CH11231. 


\section*{References}

\begin{thebibliography}{10}
\expandafter\ifx\csname url\endcsname\relax
  \def\url#1{\texttt{#1}}\fi
\expandafter\ifx\csname urlprefix\endcsname\relax\def\urlprefix{URL }\fi
\expandafter\ifx\csname href\endcsname\relax
  \def\href#1#2{#2} \def\path#1{#1}\fi

\bibitem{Chaly1997}
A.~Chaly, A.~Logatchev, S.~Shkol'nik, {Cathode spot dynamics on pure metals and
  composite materials in high-current vacuum arcs}, IEEE Trans. Plasma Sci.
  25~(4) (1997) 564--570.
\newblock \href {http://dx.doi.org/10.1109/27.640666}
  {\path{doi:10.1109/27.640666}}.

\bibitem{Almeida2013}
N.~A. Almeida, M.~S. Benilov, L.~G. Benilova, W.~Hartmann, N.~Wenzel,
  {Near-Cathode Plasma Layer on CuCr Contacts of Vacuum Arcs}, IEEE Trans.
  Plasma Sci. 41~(8) (2013) 1938--1949.
\newblock \href {http://dx.doi.org/10.1109/TPS.2013.2260832}
  {\path{doi:10.1109/TPS.2013.2260832}}.

\bibitem{Benilov2013}
M.~S. Benilov, M.~D. Cunha, W.~Hartmann, S.~Kosse, A.~Lawall, N.~Wenzel,
  {Space-Resolved Modeling of Stationary Spots on Copper Vacuum Arc Cathodes
  and on Composite CuCr Cathodes With Large Grains}, IEEE Trans. Plasma Sci.
  41~(8) (2013) 1950--1958.
\newblock \href {http://dx.doi.org/10.1109/TPS.2013.2263255}
  {\path{doi:10.1109/TPS.2013.2263255}}.

\bibitem{Barengolts2003}
S.~A. Barengolts, G.~A. Mesyats, D.~L. Shmelev, {Structure and time behavior of
  vacuum arc cathode spots}, IEEE Trans. Plasma Sci. 31~(5) (2003) 809--816.
\newblock \href {http://dx.doi.org/10.1109/TPS.2003.818449}
  {\path{doi:10.1109/TPS.2003.818449}}.

\bibitem{Mesyats2013}
G.~A. Mesyats, {Ecton Mechanism of the Cathode Spot Phenomena in a Vacuum Arc},
  IEEE Trans. Plasma Sci. 41~(4) (2013) 676--694.
\newblock \href {http://dx.doi.org/10.1109/TPS.2013.2247064}
  {\path{doi:10.1109/TPS.2013.2247064}}.

\bibitem{Anders2005d}
A.~Anders, {The fractal nature of vacuum arc cathode spots}, IEEE Trans. Plasma
  Sci. 33~(5) (2005) 1456--1464.
\newblock \href {http://dx.doi.org/10.1109/TPS.2005.856488}
  {\path{doi:10.1109/TPS.2005.856488}}.

\bibitem{Sasaki1989}
J.~Sasaki, I.~Brown, {Ion spectra of vacuum arc plasma with compound and alloy
  cathodes}, J. Appl. Phys. 66~(11) (1989) 5198--5203.
\newblock \href {http://dx.doi.org/10.1063/1.343756}
  {\path{doi:10.1063/1.343756}}.

\bibitem{Eizner1996}
B.~Eizner, G.~Markov, A.~Minevich, {Deposition stages and applications of CAE
  multicomponent coatings}, Surf. Coat. Technol. 79~(1-3) (1996) 178--191.
\newblock \href {http://dx.doi.org/10.1016/0257-8972(95)02427-1}
  {\path{doi:10.1016/0257-8972(95)02427-1}}.

\bibitem{Sasaki1993}
J.~Sasaki, K.~Sugiyama, X.~Yao, I.~G. Brown, {Multiple-species ion beams from
  titanium-hafnium alloy cathodes in vacuum arc plasmas}, J. Appl. Phys.
  73~(11) (1993) 7184--7187.
\newblock \href {http://dx.doi.org/10.1063/1.352390}
  {\path{doi:10.1063/1.352390}}.

\bibitem{Schulke1999}
T.~Sch\"{u}lke, A.~Anders, {Ion charge state distributions of alloy-cathode
  vacuum arc plasmas}, IEEE Trans. Plasma Sci. 27~(4) (1999) 911--914.
\newblock \href {http://dx.doi.org/10.1109/27.782259}
  {\path{doi:10.1109/27.782259}}.

\bibitem{Savkin2010}
K.~P. Savkin, G.~Y. Yushkov, A.~G. Nikolaev, E.~M. Oks, {Generation of
  multicomponent ion beams by a vacuum arc ion source with compound cathode},
  Rev. Sci. Instrum. 81~(2) (2010) 02A501.
\newblock \href {http://dx.doi.org/10.1063/1.3257703}
  {\path{doi:10.1063/1.3257703}}.

\bibitem{Adonin2012}
A.~Adonin, R.~Hollinger, {Development of high current Bi and Au beams for the
  synchrotron operation at the GSI accelerator facility.}, Rev. Sci. Instrum.
  83~(2) (2012) 02A505.
\newblock \href {http://dx.doi.org/10.1063/1.3670743}
  {\path{doi:10.1063/1.3670743}}.

\bibitem{Bilek1998}
M.~M.~M. Bilek, P.~J. Martin, D.~R. McKenzie, {Influence of gas pressure and
  cathode composition on ion energy distributions in filtered cathodic vacuum
  arcs}, J. Appl. Phys. 83~(6) (1998) 2965--2970.
\newblock \href {http://dx.doi.org/10.1063/1.367052}
  {\path{doi:10.1063/1.367052}}.

\bibitem{Eriksson2013}
A.~O. Eriksson, I.~Zhirkov, M.~Dahlqvist, J.~Jensen, L.~Hultman, J.~Rosen,
  {Characterization of plasma chemistry and ion energy in cathodic arc plasma
  from Ti-Si cathodes of different compositions}, J. Appl. Phys. 113~(16)
  (2013) 163304.
\newblock \href {http://dx.doi.org/10.1063/1.4802433}
  {\path{doi:10.1063/1.4802433}}.

\bibitem{Zhirkov2013}
I.~Zhirkov, A.~O. Eriksson, J.~Ros\'{e}n, {Ion velocities in direct current arc
  plasma generated from compound cathodes}, J. Appl. Phys. 114 (2013) 213302.
\newblock \href {http://dx.doi.org/10.1063/1.4841135}
  {\path{doi:10.1063/1.4841135}}.

\bibitem{Zhirkov2014}
I.~Zhirkov, A.~O. Eriksson, A.~Petruhins, M.~Dahlqvist, A.~S. Ingason,
  J.~Ros\'{e}n, {Effect of Ti-Al cathode composition on plasma generation and
  plasma transport in direct current vacuum arc}, J. Appl. Phys. 115~(12)
  (2014) 123301.
\newblock \href {http://dx.doi.org/10.1063/1.4869199}
  {\path{doi:10.1063/1.4869199}}.

\bibitem{Nikolaev2013}
A.~G. Nikolaev, G.~Y. Yushkov, K.~P. Savkin, E.~M. Oks, {Angular Distribution
  of Ions in a Vacuum Arc Plasma With Single-Element and Composite Cathodes},
  IEEE Trans. Plasma Sci. 41~(8) (2013) 1923--1928.
\newblock \href {http://dx.doi.org/10.1109/TPS.2012.2236363}
  {\path{doi:10.1109/TPS.2012.2236363}}.

\bibitem{Nikolaev2014}
A.~G. Nikolaev, K.~P. Savkin, G.~Y. Yushkov, E.~M. Oks, {Ion angular
  distribution in plasma of vacuum arc ion source with composite cathode and
  elevated gas pressure}, Rev. Sci. Instrum. 85 (2014) 02B501.
\newblock \href {http://dx.doi.org/10.1063/1.4824641}
  {\path{doi:10.1063/1.4824641}}.

\bibitem{Franz2013a}
R.~Franz, P.~Polcik, A.~Anders, {Ion Charge State Distributions of Al and Cr in
  Cathodic Arc Plasmas From Composite Cathodes in Vacuum, Argon, Nitrogen, and
  Oxygen}, IEEE Trans. Plasma Sci. 41~(8) (2013) 1929--1937.
\newblock \href {http://dx.doi.org/10.1109/TPS.2013.2254135}
  {\path{doi:10.1109/TPS.2013.2254135}}.

\bibitem{Anders2008}
A.~Anders, {Cathodic Arcs - From Fractal Spots to Energetic Condensation},
  Springer, New York, USA, 2008.

\bibitem{Oks1995}
E.~Oks, I.~Brown, M.~Dickinson, R.~MacGill, H.~Emig, P.~Sp\"{a}dtke, B.~H.
  Wolf, {Elevated ion charge states in vacuum arc plasmas in a magnetic field},
  Appl. Phys. Lett. 67~(2) (1995) 200--202.
\newblock \href {http://dx.doi.org/10.1063/1.114666}
  {\path{doi:10.1063/1.114666}}.

\bibitem{Oks1996b}
E.~Oks, G.~Y. Yushkov, {Some features of vacuum arc plasmas with increasing gas
  pressure in the discharge gap}, in: Proc. 17th Int. Symp. Discharges Electr.
  Insul. Vac., Vol.~2, Berkeley (CA), USA, 1996, pp. 584--588.
\newblock \href {http://dx.doi.org/10.1109/DEIV.1996.545430}
  {\path{doi:10.1109/DEIV.1996.545430}}.

\bibitem{Yushkov2000a}
G.~Y. Yushkov, A.~Anders, E.~M. Oks, I.~G. Brown, {Ion velocities in vacuum arc
  plasmas}, J. Appl. Phys. 88~(10) (2000) 5618.
\newblock \href {http://dx.doi.org/10.1063/1.1321789}
  {\path{doi:10.1063/1.1321789}}.

\bibitem{Anders2001a}
A.~Anders, B.~Yotsombat, R.~Binder, {Correlation between cathode properties,
  burning voltage, and plasma parameters of vacuum arcs}, J. Appl. Phys.
  89~(12) (2001) 7764--7771.
\newblock \href {http://dx.doi.org/10.1063/1.1371276}
  {\path{doi:10.1063/1.1371276}}.

\bibitem{Mraz2006a}
S.~Mr\'{a}z, J.~M. Schneider, {Influence of the negative oxygen ions on the
  structure evolution of transition metal oxide thin films}, J. Appl. Phys.
  100~(2) (2006) 023503.
\newblock \href {http://dx.doi.org/10.1063/1.2216354}
  {\path{doi:10.1063/1.2216354}}.

\end{thebibliography}
\end{document}